# Bayesian approach to the PC component

Jie Yao[1,] *, Zhang Kai[2], Eric Rose[1], Edward Valachovic[1]

1. Department of Epidemiology and Biostatistics, College of Integrated Health Sciences, University at Albany, State University of New York, Rensselaer, NY, 12144, USA

2. Department of Population and Community Health, College of Public Health, The University of North Texas Health Science Center at Fort Worth, Fort Worth, TX, 76107, USA

* Correspondence: jyao@albany.edu (J.Y.)

**Abstract**

Time series with multiple periodically correlated (MPC) components is a complex problem with comparatively limited prior research. Most existing time series models are designed to accommodate simple periodically correlated (PC) components and tend to be sensitive to over-parameterization and optimization issues and are also unable to model complex PC components patterns in a time series. Frequency separation techniques can be used to maintain the correlation structure of each specific PC component, whereas Bayesian techniques can combine new and existing prior information to update beliefs about these components. This study introduces a method to combine the frequency separation techniques and Bayesian techniques to forecast PC and MPC time series data in a two-stage form, which is expected to show the new method's suitability in modeling MPC components compared to classical methods.

**Introduction**

Time series data are observations obtained through repeated measurements over time. Time series are often composed of several interacting factors such as trends, periodic patterns such as seasonality, and random variation or noise. Many time series exhibit multiple periodically correlated (MPC) patterns. Periodic components with a given period, $p$, exhibit strong correlations between data points that are $kp$ time points removed, or lagged, where $k$ is an integer multiple. For example, the number of retail banking call arrivals recorded every 5-minute interval from 7:00 a.m.

to 9:05 p.m. on weekdays follows a daily periodic pattern with a period of (14 1/12) hours × 12 observations / hour = 169 observations and a weekly periodic pattern with a period of 169 observations/day × 5 days/week = 845 minutes.[1] An extended version of this series may also show an annual periodic or seasonal pattern. Similar multiple periodic patterns can be observed in areas such as daily hospital admissions, ATM cash withdrawal requests, electricity and water consumption, and website access.[1] Forecasting time series with MPC components, where periodical cycles vary in length, is often challenging. Therefore, identifying and accounting for PC component effects are essential for accurate time series forecasting.

Among various current forecasting methods, the seasonal autoregressive integrated moving average (ARIMA) model and exponential smoothing technique are considered classical approaches.[2] However, their basic forms are designed for single PC component modeling and can have difficulty handling MPC patterns. Numerous studies have sought to extend these traditional statistical forecasting models to accommodate MPC components.[1,2,3] Notable approaches include the double seasonal ARIMA model, an adaptation of the exponential smoothing technique from the simple Holt-Winters method, and the hidden Markov model with multiple seasonality. Despite these advancements, existing methods often struggle with over-parameterization and optimization challenges and fail to effectively capture complex seasonal patterns in time series data.

MPC components operate at distinct frequencies, corresponding to the reciprocals of their respective periods. Although signals at these frequencies may overlap in the time domain, they remain uncorrelated, an advantageous fact taken from mathematics and see in many real world examples in physics and other fields, with each frequency behaving independently.[4] By isolating and filtering an MPC time series based on its spectral density at individual PC component frequencies, it is possible to extract a set of PC component time series, each exhibiting a single PC structure.[5] For example, as shown in Figure 1, the MPC wave *2sinx+3sin2x* can be separated into two PC components: *2sinx and 3sin2x*.

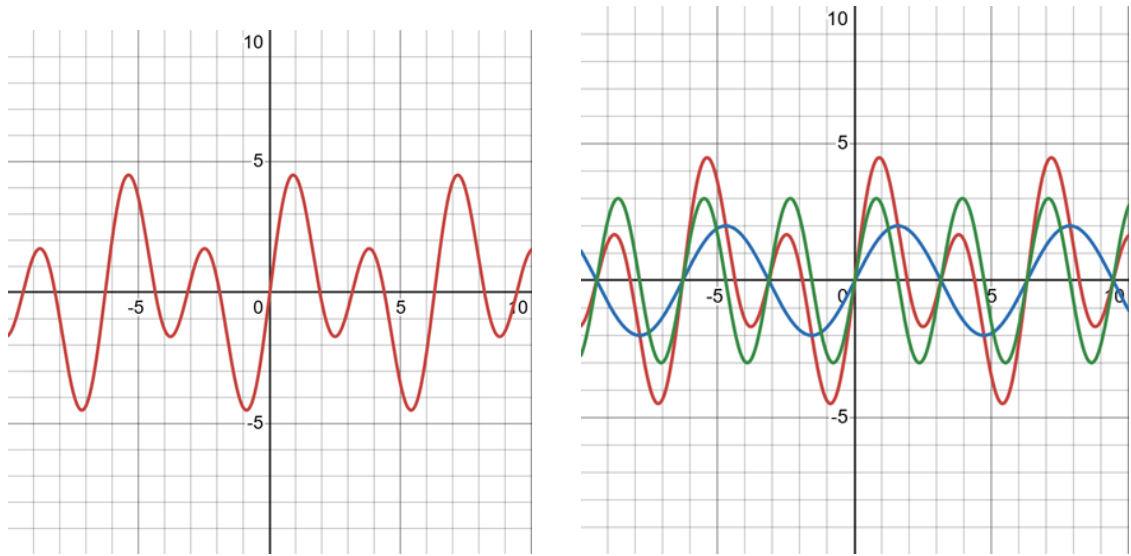

Figure 1: The sine waves of *2sinx+3sin2x*

This study uses the novel VBPBB method, which applies a bandpass filter centered on the frequency of a periodically correlated (PC) component, preserving time series variations occurring at or near the targeted frequency.[5] In general, block bootstrapping has difficulty reproducing the correlation structure of PC time series. First, arbitrarily bootstrapping with block lengths less than the given period $p$, will sever these correlations. By isolating and filtering MPC components based on their spectral density at specific PC frequencies, a set of PC time series is generated by VBPBB, each exhibiting a distinct PC structure. Each of these PC component time series, characterized by a single PC structure, can then undergo individual block bootstrapping with an appropriately chosen block size to preserve its unique correlation structure.[5] Compared to other periodic block bootstrap (PBB) techniques which cannot preserve correlation structure, VBPBB yields smaller confidence interval (CI) sizes for the periodic mean, enhancing estimation precision. It may also be beneficial to VBPBB even in the case of a single PC time series, as it will separate the PC component from other potential components, even those that are not PC such as noise. This may improve the investigation of the properties of the component itself.

Many models utilize Bayesian techniques, which integrate the likelihood function—representing observed data—with prior knowledge about an unknown statistical parameter to update prior information and construct the final Bayesian beliefs, known as posterior distributions.[6] The Bayesian technique has the disadvantage of requiring the placement of prior beliefs on the unknown parameters but provides the advantage of producing posterior distributions over the classical approach. Bayesian model averaging helps enhance forecast accuracy because it balances different perspectives from multiple models, accounts for uncertainty, and reduces the risk of relying on a single potential model. When prior distributions on multiple parameters are independent, the resulting posterior distributions remain independent, allowing Bayesian inference to be conducted separately for each parameter. For parameter estimation, Markov Chain Monte Carlo (MCMC) algorithms are employed. These methods enable sampling from a posterior distribution for the probability of parameter $\theta$ given that data $D$ without requiring its explicit analytic form. The MCMC process ultimately yields a set of parameter vectors $\theta$ with density facilitating parameter estimation.[7] The Metropolis–Hastings algorithm is a MCMC method designed to generate a sequence of random samples from a probability distribution when direct sampling is challenging. This process involves two steps: first, a new sample is proposed based on the previous one, and then the proposed sample is either accepted and added to the sequence or rejected based on the probability distribution's value at that point. The generated sequence can be utilized to approximate the target distribution or estimate integrals. More complex models can be developed by layering multiple simple stages within hierarchical Bayesian frameworks.[8]

Tongkhow and Kantanantha introduced Bayesian forecasting models by modifying Yelland's approach, altering the treatment of outliers, autoregression, and certain prior distributions.[9, 10]

Our methods differ from forecasting techniques that handle large numbers of parameters in each data point. We do not directly model the distribution of our data points, as would be the case in a dynamic factor model (e.g. Forni and Reichlin, 1998; Forni et al., 2000). For a PC component, estimating the parameters at every data point over a period of time would make the analysis overly complex. Instead of analyzing the parameters at each data point, estimating the amplitude of a PC component can reduce the calculation. The amplitude of a PC component is a measure of its change in a single period which is not directly measurable. Use the existing information of the period can with most observations to transform each into updating the information about the amplitude. An alternate approach is to estimate the amplitude of each PC component which is the only unknown parameters when frequency is pre-determined since each PC component is defined by the amplitude and frequency. This is illustrated in the following figure 2. MPC components are the combination of each filtered PC component. With the help of VBPBB, we can filter MPC components into different significant PC components. Thus, we can perform a Bayesian analysis on each PC component. With PC components operating on different frequencies, the joint prior distributions on multiple parameters are independent, then the posterior distributions are also independent. Independence allows us to more easily specify the prior, when amplitude is uncorrelated. The independence of unknown parameters is assumed here.

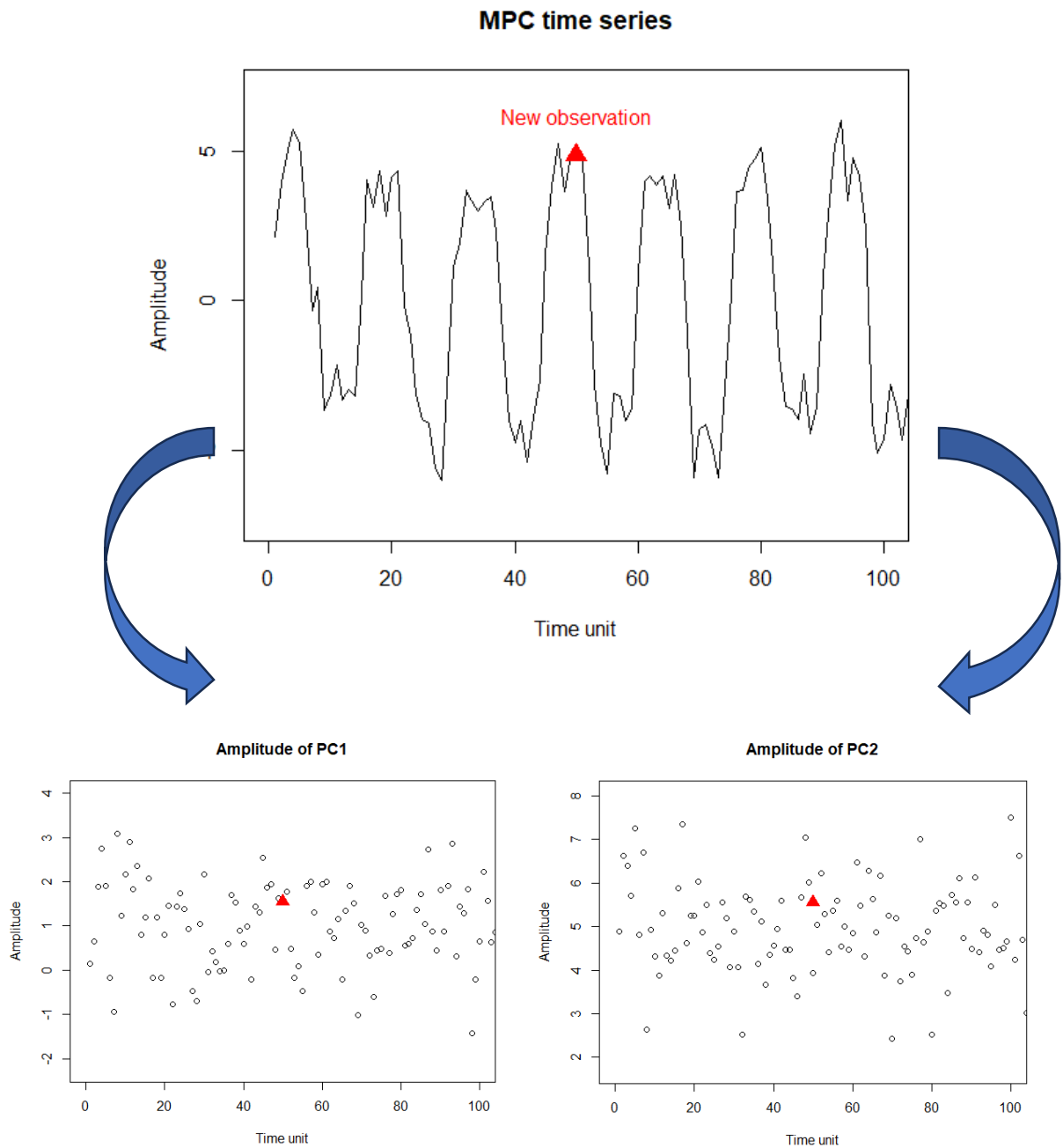

Figure 2. Illustration of a new observation in MPC time series updating the information of amplitude of both PC1 and PC2.

This article proposes a two-stage method for addressing time series exhibiting MPC components, which eliminates the need for pre-determined PC components periods. In the first stage, we use

VBPBB to filter the MPC time series to different significant single PC components of the corresponding frequency and its harmonics, allowing the preservation of time series variation occurring at or near the targeted component frequency. In the second stage, for each separated PC component, we employ a suitable Bayesian time series model to estimate the amplitude of each PC component for the corresponding frequency. Thus, we can sum up the fitted PC components to do the forecasting.

**Methods**

Exponential Smoothing Models

An exponential smoothing model is a technique used to minimize fluctuations in time series data by applying exponentially decreasing weights to past observations, giving greater importance to more recent data points. The double exponential smoothing model builds on this by incorporating a linear trend component, while the triple exponential smoothing model, also known as the Holt-Winters model, extends the approach further to account for both trend and seasonal patterns.[1]

SARIMA Model

The Seasonal Autoregressive Integrated Moving Average (SARIMA) model extends the ARIMA model to address time series data that includes both trends and seasonal patterns. By introducing additional parameters, SARIMA effectively accounts for periodic fluctuations. It integrates non-seasonal components with seasonal terms, enabling it to capture both short-term variations and recurring seasonal trends. This model is especially useful for forecasting data with consistent seasonal cycles, such as monthly sales, quarterly financial metrics, or yearly climate patterns. This seasonal component of the ARIMA model is denoted by capital letters, SARIMA *(p, d, q) (P, D, Q)m,* where the last bracket indicates the seasonal factor parameters for the order of autoregressive, integration and moving average parts of the model. The first bracket indicates the non-seasonal parameters.[11]

VBPBB

The VBPBB utilizes a bandpass filter to isolate and process the PC time series, permitting only a narrow range of frequencies centered around the corresponding frequency of the PC component to pass through. Frequency outside this range, known as the stopband, is suppressed. The bandpass filter applied in this process is the Kolmogorov-Zurbenko Fourier Transform (KZFT) filter which is a symmetric band pass filter around a predetermined frequency. KZFT filters are capable of selectively isolating specific parts of the frequency domain to eliminate interfering frequencies.[4]

After PC components separation, VBPBB then block bootstraps each individual PC component time series. Using methods outlined earlier, such as using block lengths equal to the period of a particular PC component, VBPBB will preserve each PC structure. This allows the investigation of PC components individually in the MPC time series.

Finally, to bootstrap the complete MPC time series or any selected set of PC components, each PC component is individually block bootstrapped using an appropriate block length. The resulting resamples from each component are then summed to form a single resample representing the intended set of components. By generating an equal number of resamples for each PC component, the total number of resamples will match that of the desired set of components, including the full MPC time series.

Bayesian methods

Bayesian methods are a class of statistical techniques based on Bayes' theorem, which describes how to update the probability of a hypothesis as new data becomes available. These methods combine prior knowledge or beliefs (expressed as a prior distribution) with observed data (likelihood) to generate a posterior distribution that reflects updated beliefs about the parameters of interest. A hierarchical Bayesian model (Congdon, 2010) is formulated as Equation:

$$p(\theta|D) = \frac{p(D|\theta)p(\theta)}{p(D)}$$

where $p(D|\theta)$ represents the probability of event $D$ occurring given that parameter $\theta$ is true, which can also be interpreted as the likelihood, $p(\theta|D)$ is a posterior distribution which stands for the probability of parameter θ given that data D is true. , $p(\theta)$ is a prior distribution of $\theta$, which summarizes any priori or alternative knowledge on the distribution of θ and *p(D)* is the marginal distribution of data *D*, which represents the probabilities of data D respectively without any given conditions.[10]

The MCMC algorithms are used for parameter estimation. The MCMC methods provide a way to sample from $p(\theta|D)$ without necessarily knowing its analytic form. The final result of MCMC is a set of vectors $\theta$ with density $p(\theta|D)$ in which the model parameters can be estimated. In this study, the Metropolis–Hastings algorithm is used to generate a series of random samples from a probability distribution when direct sampling is difficult. MCMC uses a Markov Chain that eventually "forgets" where it started, and samples from the true posterior. Over time, the chain approaches equilibrium, and from that point on, the samples are like drawing from the posterior directly. It operates in two steps: first, a candidate sample is proposed based on the previous one, and then it is either accepted and included in the sequence or rejected, depending on the probability ratio between the proposed sample and the previous sample. MCMC accepts the proposed value with probability ratio greater than 1; otherwise, it stays at the current value. This resulting sequence can be employed to approximate the target distribution or computing integrals underneath the distribution. The most common hierarchical Bayesian model has three stages. A distribution for the data given parameters is specified at the first stage, prior distributions for parameters given hyper-parameters are specified at the second stage and the distribution for hyper-parameters are specified at the third stage. Complicated models can be built through the specification of several simple stages under hierarchical Bayesian models.

The Proposed VBPBB - Bayesian Model

This model was developed by combining the principles of the Bayesian technique and VBPBB. The design was inspired by the observation that, compared to the complexity and difficulty of modeling MPC data, modeling data at a single PC component level for corresponding frequency is much more feasible. Therefore, by individually modeling each significant PC component within the MPC data by filtering out the noise and subsequently integrating them, we can effectively construct a comprehensive model of the MPC data. In the first stage, the VBPBB employs a bandpass filter to segregate and filter the PC time series for different period $p$, allowing only a narrow band of frequencies around the corresponding frequency of the PC component to pass through. VBPBB can selectively resample a PC component time series to maintain the correlation structure of that specific PC component. This is done without resampling other unrelated components, such as noise or a linear trend, which could unnecessarily increase bootstrap variability. Thus, only those significant PC components that construct the MPC data will be filtered out in this step. By summing up all the fitted significant PC components, the VBPBB-Bayesian model can generate the model for MPC data. In this case, to selectively pass or retain only one PC component in each KZFT bandpass filter, the width of passed frequencies should be set to no more than halfway between the minimum bandwidth among all frequencies to be filtered so that we can get the significant PC component for the corresponding frequency.

After generating significant single PC components, we need to build a model that can accurately model the filtered PC component data. However, the most classical methods just typically assume a distribution for each data point, resulting in one parameter being set for each point within a cycle. This means that the longer the cycle, the more parameters are required, leading to extremely high computational demands when modeling long-period data. In contrast, our model aims to remain as simple as possible; therefore, we do not perform detailed analysis for every data point within a cycle. In this model, our model will just model the amplitude instead of all the parameters for each data point within the corresponding cycle.

The theorem that we can only estimate the amplitude of the significant PC component is based on (periodogram). For each PC component, it can be expressed as a sine wave with differing frequencies (how long it takes to complete a full cycle) and amplitudes (maximum/minimum value during the cycle). This fact can be utilized to examine periodic (cyclical) behavior in a time series. Imagine constructing a single sine wave as a time series observed in discrete time. Suppose that we write this sine wave as $x_t$, a function in time *t*:

$$x_t = A sin(2\pi \omega t + \varphi)$$

*A* is the amplitude which determines the maximum absolute height of the curve. $\omega$ is the frequency that controls how rapidly the curve oscillates. $\varphi$ is the phase that determines the starting point, in angle radians, for the sine wave. To temporarily simplify things, suppose that $\varphi = 0$ and think about the quantity $2\pi\omega t$. Assume p = number of time points/observations for a full cycle and that *p = 1/T*. As we move through time from *t* = 0 to *t = p*, the value of $2\pi\omega t$ is $2\pi \frac{1}{p} t$ ranges from 0 to $2\pi$. Thus, in our model, we aim to estimate the unknown parameter of amplitude as the significant PC components identified by VBPBB provide the existing information about the amplitude.

In the second stage, based on the filterer significant PC components from the first stage, we will separately model the significant PC components through Bayesian technique. Use the existing information of the period can with most observations to transform each into updating the information about the amplitude. For each PC component, let $Y_t$ be a filtered PC time series data at time $t$, $t = 1, ..., n$. $Y_t$ is assumed normally distributed whose mean can detect trend, PC component and account for some covariates. The proposed model is defined as:

$$Y_t \sim N(A\sin\left(2\pi\frac{1}{p}t\right), \sigma^2)$$

where $A$ is the amplitude, $p$ is the predetermined period, and $\sigma^2$ is the common variance of $Y_t$. Therefore, $A$ and $\sigma^2$ are the parameters we aim to estimate in this stage.

For parameter estimation, the MCMC algorithm called Metropolis–Hastings algorithm is used. It is particularly useful for Bayesian inference and complex models where the posterior distribution cannot be sampled directly. We are seeking the joint posterior distribution of the unknown parameters $A$ and $\sigma^2$ from a normal distribution according to the PC components generated from VBPBB. Then, if we write the Bayes rule, we have the following expression:

$$p(A, \sigma|Y_t) = \frac{p(Y_t|A, \sigma)p(A, \sigma)}{\iint p(Y_t|A, \sigma)p(A, \sigma)p(A)p(\sigma)}$$

Omitting the denominator that is a constant and using the proportionality symbol:

$$p(A, \sigma|Y_t) \propto p(Y_t|A, \sigma)p(A, \sigma)$$

If $A$ and $\sigma^2$ are independent, we can re-rewrite the joint prior as two independent priors:

$$p(A, \sigma) = p(A)p(\sigma^2)$$

As both parameters are ranging from 0 to infinite, we can first assume a gamma distribution as prior for both parameters $A$ and $\sigma^2$. Thus, the prior distributions for Bayesian methods are assigned to each parameter as follows:

$$p(A) \sim \text{Gamma}(1, 0.1)$$
$$p(\sigma^2) \sim \text{Gamma}(1, 0.0001)$$

The algorithm constructs a Markov chain that converges with the desired target distribution. The whole process is illustrated in the following figure 3.

Here's a step-by-step outline of the algorithm:

The proposal step:
Generate a random candidate state: $x'$ according to $g(x'|x_t)$.

The accept-reject step:
Calculate the acceptance probability: $A(x', x_t) = \min\left(1, \frac{P(x')}{P(x_t)} \frac{g(x_t|x')}{g(x'|x_t)}\right)$.

Accept or reject:
1. Generate a uniform random number $u \in [0, 1]$.
2. If $u \leq A(x', x_t)$, then accept the new state and set $x_{t+1} = x'$.
3. If $u > A(x', x_t)$, then reject the new state, and copy the old state forward $x_{t+1} = x_t$.

Convergence:
Repeat the process for a sufficient number of iterations to ensure convergence to the target distribution. Convergence in MCMC refers to the point at which the Markov chain has run long enough that the samples it produces are effectively representative of the target distribution (often a posterior distribution in Bayesian statistics). Before convergence, the samples are biased by the initial starting values; after convergence, they reflect the true distribution. MCMC chains start out biased by the initial value. It takes time to reach the stationary distribution (i.e., sampling from the actual posterior). Thus, discarding burn-in ensures your final samples are representative and reliable. The samples obtained after the burn-in period (initial discarded samples) can be used to approximate expectations, distributions, or model parameters.

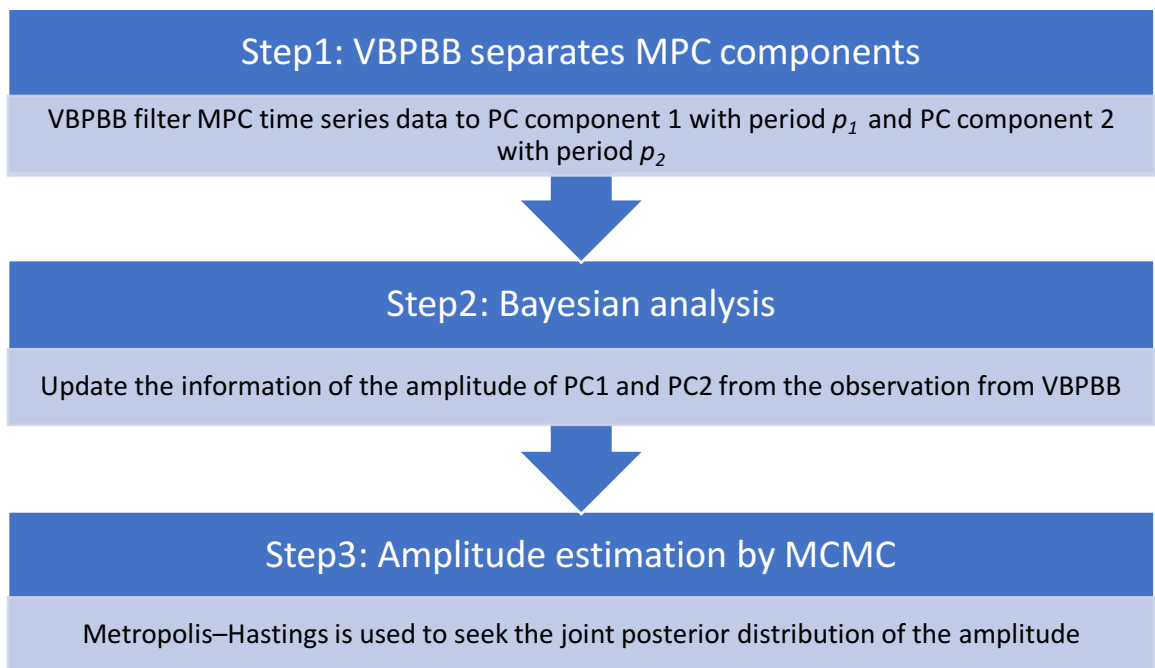

Figure 3. The flowchart of the VBPBB – Bayesian model step by step.

**Simulation**

This simulation illustrates the application of the proposed VBPBB - Bayesian model and its results by simulating a time series and applying VBPBB - Bayesian model. Periodograms, as explained by Wei (1990).[12] visually depict the spectral energy across different frequencies in a time series, enabling the observation of the distinct impacts of VBPBB - Bayesian model on the original data.

Analysis is performed in R version 4.1.1 (2013) statistical software using the KZFT function in the KZA package, see Close and Zurbenko (2013) for more detail,[13] with datasets as a time series measured on an ordered interval dimension, in this case time. All time series are constructed with 300-time units. First, two periodically correlated sine wave signals with different periods, or frequencies, where the time coordinate determines the phase of the sine waves, are summed. The result is a multiple periodically correlated time series of interacting waves entangled in the time domain. Next, random variation is introduced by generating equal length vectors of elements randomly selected from a standard normal distribution. These random variations are then combined with the summed PC components. The final MPC time series of data is composed of the two PC components obscured by noise, seen in the figure below, and this would represent the data ordinarily available at the time of analysis.

In these simulations, PC1 has period $p1 = 15$ with amplitude=5 and PC2 has period $p2 = 50$ with amplitude =50, and noise has the standard deviation of 100 which were presented in figure 4 and 5. The simulation performs VBPBB by separating the PC components using KZFT filters, and block bootstrapping each component according to the described strategy, with fixed block size 15 for PC1, and 50 for PC2. In this scenario, KZFT filters are centered above the PC component frequencies, while choosing parameters to exclude the other PC component outside of the cut-off boundary for that filter. Then we applied Metropolis–Hastings algorithm on those filtered data. For M-H algorithm, we first assume a simple gamma distribution as priors for amplitude and variance with $p(A) \sim gamma(1, 0.1)$ and $p(\sigma^2) \sim gamma(1, 0.0001)$. For the proposal values for $A$ we will use a normal distribution with mean equivalent to the value of $A$ in the previous step of the chain and standard variation 2 and for the standard deviation σ we will use the previous value of variance in the chain added to a random value drawn from a uniform distribution between -1/2 and 1/2. We set the initial values 2 and 6 for $A$ and $\sigma$ and do 3000 iterations for each simulation.

For simulation of single PC component time series data $X_t = 5\sin\left(\frac{2\pi}{15\,t}\right) + \varepsilon_t$, where $\varepsilon_t \sim N(0, 100)$. For fitted model, the KZFT $m$=6, f=1/30 filtered time series of single PC. After discarding the initial 300 values of the iteration chain, we got the mean of amplitude at 4.96 with the variance of where the true amplitude is 5. For simulation of double PC component time series data $X_t = 5\sin\left(\frac{2\pi}{15\,t}\right) + 10\sin\left(\frac{2\pi}{50\,t}\right) + \varepsilon_t$, where $\varepsilon_t \sim N(0, 100)$. For fitted model, the KZFT $m$ = 6, f = 1/30 filtered time series of PC1 and the KZFT $m$ = 7, f = 1/50 filtered time series of PC2. After discarding the initial 300 values of the iteration chain, we got the mean of amplitude of the iteration chain at 5.03 for PC1 and mean of amplitude of the iteration chain at 10.36 for PC2.

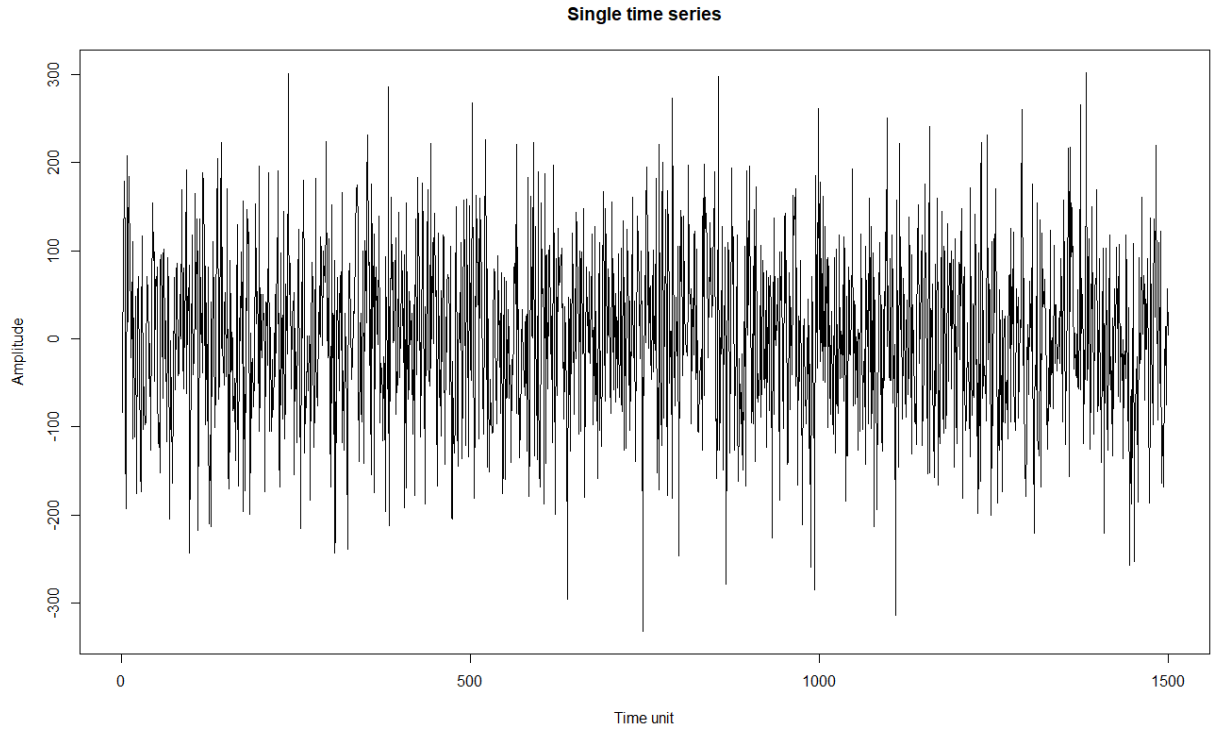

Figure 4. Sigle PC time series data of $X_t = 5sin\left(\frac{2\pi}{15\,t}\right) + \varepsilon_t$.

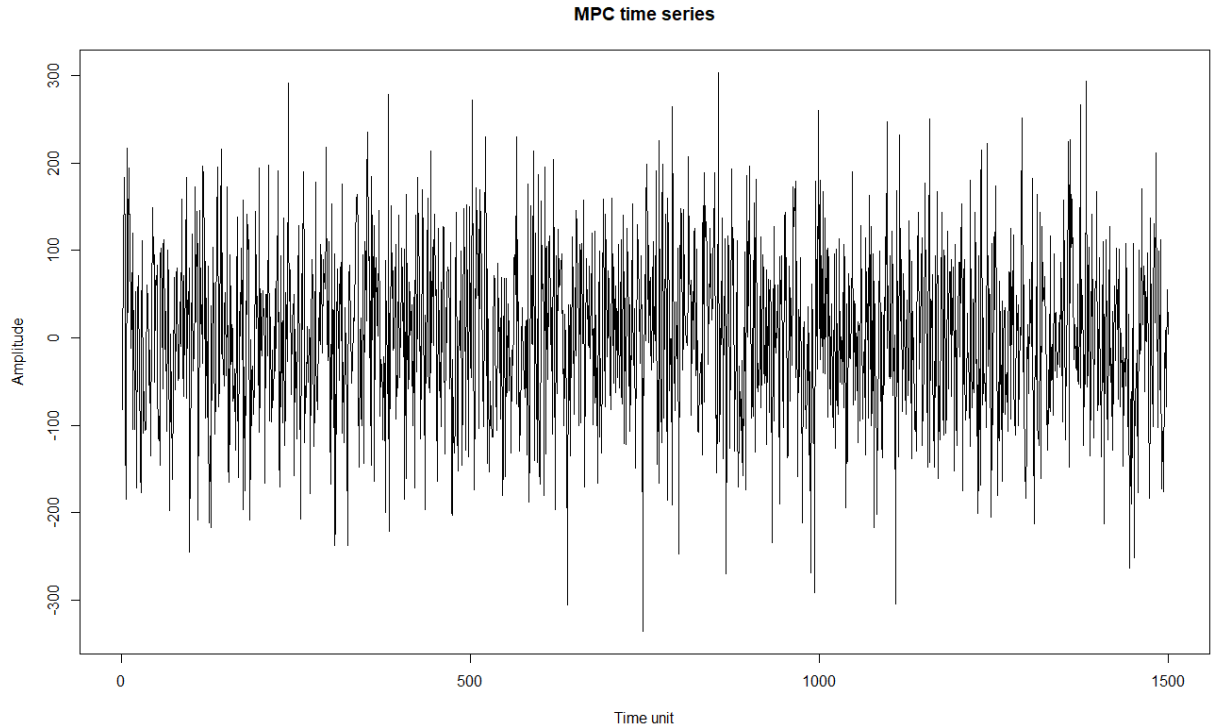

Figure 5. Double PC time series data of $X_t = 5sin\left(\frac{2\pi}{15\,t}\right) + 10sin\left(\frac{2\pi}{50\,t}\right) + \varepsilon_t$.

**Real data applications**

The practical use for researchers and potential advantages of the VBPBB-Bayesian model can be demonstrated in real data applications. In this data application, we take U.S. monthly milk production from January 1962 through December of 1975, the detrending data can be seen in Figure 4. Data were recorded monthly, enabling investigation of monthly frequency, in addition to their harmonics, based on the periodograms of components. The time series has 168 observations. Analysis is performed in R version 4.1.1 (2013) statistical software.

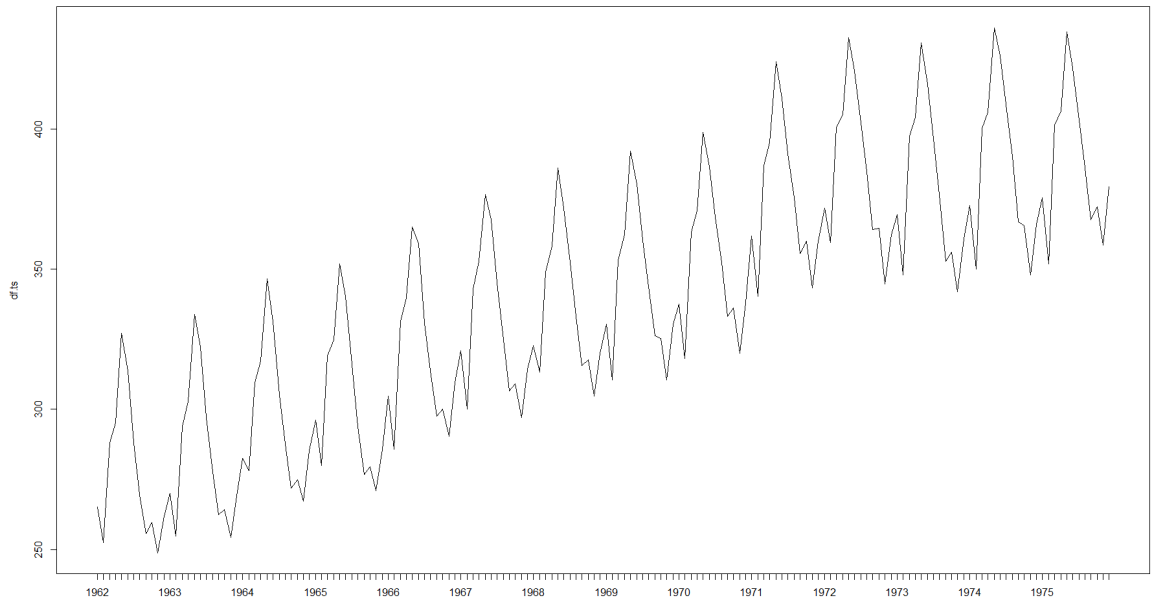

Figure4. U.S. monthly milk production from January 1962 through December of 1975.

Firstly, we will detrend the time series data since there is a linear trend. Figure 5 shows the U.S. monthly milk production time series in black. For fitted model, the $KZFT m=3$, f=1/12 filtered time series of PC1 and the $KZFT m=3$, f=2/12 filtered time series of PC2. After the VBPBB-Bayesian model fitting, we got the estimated mean of 17.63 and standard deviation of 9.92 for the amplitude of PC1 and the estimated mean of 5.43 and standard deviation of 6.21 for the amplitude of PC2. By applying the parameters of PC1 and PC2, the fitted time series was in red in the figure 5. This example demonstrates the suitability of VBPBB-Bayesian model in MPC time series data.

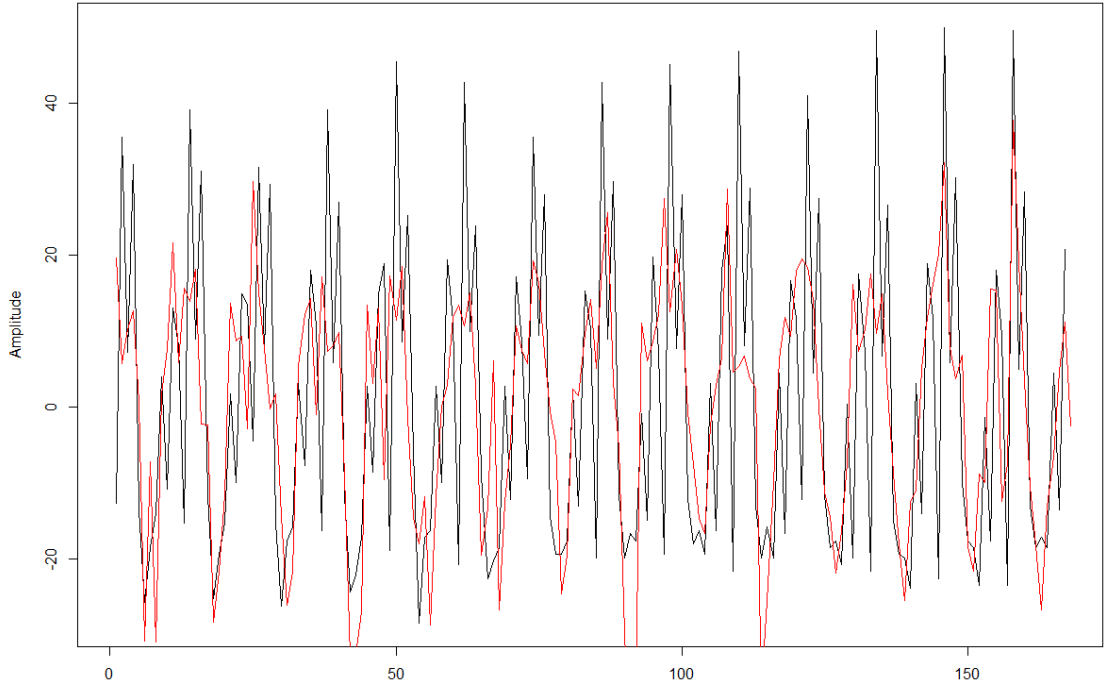

Figure5.

**Discussion**

The VBPBB-Bayesian model is a useful tool in time series analysis to forecast the MPC data by summing up the significant PC components for corresponding frequencies as well as it's harmonic frequencies. The VBPBB-Bayesian model identified significant PC components in the MPC data through VBPBB. VBPBB separates the MPC time series into significant PCs for a specific frequency and its harmonics, which preserves variations in the time series that occur at or near the targeted frequency. Utilizing Bayesian techniques, the model updates information by incorporating prior knowledge of amplitude, ultimately yielding a final amplitude estimation for each significant

PC component. Consequently, the fitted MPC time series can be obtained by summing the fitted significant PC components, eliminating the need to analyze parameters at each data point.

Our methods have several strengths. Firstly, the VBPBB bootstrap approach selectively resamples a PC component time series to preserve its correlation structure, avoiding unrelated components like noise or linear trends that could add unnecessary variability. Secondly, Bayesian statistics can aid in inferring the decomposition model, particularly because of the additive nature of the equation: A time series is the sum of period and trend signals, embedded with changepoints. Finally, the only parameter we need to estimate is the amplitude, so we avoid setting up a large number of the parameters for each data point. Thus, we can reduce the dimension of our model. However, the VBPBB-Bayesian model has limitations. The performance of VBPBB is intricately linked to the selection of arguments for bandpass filtration. There are restrictions on which frequencies can be detected, and the degree of proximity two frequencies can have while remaining distinguishable through KZFT filters. The conjugate inverse gamma prior is one of the prior distributions having been suggested in Bayesian analysis, the selection of prior distribution will influence the results. Simulated data may not fully reflect the complexities and nuances of the actual data generation process.

**Conclusion**

As this study has shown, in time series analysis, unlike the classical methods which cannot handle MPC data forecasting, we proposed this innovative VBPBB-Bayesian model, a simple yet powerful forecasting procedure that combines VBPBB technique and Bayesian technique, which can forecast MPC time series data. The VBPBB-Bayesian model help preserve the variations of MPC components and use the updated information of prior knowledge of amplitude for each PC component to get the final estimation of the amplitude. Thus, the procedure has a limited number of parameters to specify. We reveal its competence of forecasting by simulations and real data applications.

**Declaration of interests**

The authors declare they have no competing financial interests.

**Acknowledgments**

This research has received no external funding.

**Author contribution**

**Jie Yao**: Conceptualization, Methodology, Data curation, Formal analysis, Visualization, Writing - Original Draft, Writing - Review & Editing. **Edward Valachovic**: Conceptualization, Methodology, Data curation, Critical review & editing, Supervision, Funding acquisition. All authors have read and approved the final draft of the paper.